%% file: main.tex
\documentclass[sigconf,screen,authorversion]{acmart}  

\AtBeginDocument{%
  \providecommand\BibTeX{{%
    \normalfont B\kern-0.5em{\scshape i\kern-0.25em b}\kern-0.8em\TeX}}}

\copyrightyear{2021}
\acmYear{2021}
\setcopyright{acmlicensed}
\acmConference[KDD '21]{Proceedings of the 27th ACM SIGKDD Conference on Knowledge Discovery and Data Mining}{August 14--18, 2021}{Virtual Event, Singapore}
\acmBooktitle{Proceedings of the 27th ACM SIGKDD Conference on Knowledge Discovery and Data Mining (KDD '21), August 14--18, 2021, Virtual Event, Singapore}
\acmPrice{15.00}
\acmDOI{10.1145/3447548.3469053}
\acmISBN{978-1-4503-8332-5/21/08}

\acmSubmissionID{ads2181a}

\usepackage{adjustbox}

\usepackage{todonotes}
\usepackage{hyperref}
\newcounter{hoifung}

\newcounter{jianfeng}

\newcounter{tristan}

\newcounter{jincli}

\newcounter{eric}
\begin{document}
\fancyhead{}

\newcommand\DNAME{Microsoft Biomedical Search}

\title[Domain-Specific Pretraining for Vertical Search: Case Study on Biomedical Literature]{Domain-Specific Pretraining for Vertical Search: \\Case Study on Biomedical Literature}



\author{Yu Wang,* Jinchao Li,* Tristan Naumann,*  Chenyan Xiong, Hao Cheng, Robert Tinn, Cliff Wong, Naoto Usuyama, Richard Rogahn, Zhihong Shen, Yang Qin, Eric Horvitz, Paul N.~Bennett, Jianfeng~Gao, Hoifung Poon}
\thanks{*These authors contributed equally to this research.}
\email{yuwan, jincli, tristan, cxiong, chehao, rotinn, clwon,naotous,}
\email{rrogahn, zhihosh, yaq, horvitz, pauben, jfgao, hoifung @microsoft.com}
\affiliation{%
  \institution{Microsoft Research}
  \streetaddress{One Microsoft Way}
  \city{Redmond}
  \state{WA}
  \postcode{98052}}

\renewcommand{\shortauthors}{Y. Wang, J. Li, T. Naumann, et al.}

\newcommand{\eat}[1]{\ignorespaces}

\input{0_abstract}




\maketitle


\input{1_introduction.tex}

\input{2_self-sup}

\input{3_covid}

\input{4_ms-biomed-search}
\input{5_discussion}
\input{6_conclusion}



\vspace{-3mm}
\begin{acks}
The authors thank 
Grace Huynh, Miah Wander, Michael Lucas, Rajesh Rao, Mu Wei, and Sam Preston for their support in assessing ranking relevance; as well as,
Mihaela Vorvoreanu, Dean Carignan, Xiaodong Liu, Adam Fourney, and Susan Dumais for contributing their expertise and shaping \DNAME.  We thank colleagues at the Cleveland Clinic Foundation for composing and sharing a sample of COVID-19--centric queries spanning a broad range of biomedical topics.
\end{acks}    
\vspace*{-\baselineskip}







\bibliographystyle{ACM-Reference-Format}
\bibliography{ref_main,ref_datasets,ref_software,ref_tristan}










\end{document}

%% file: 0_abstract.tex
\begin{abstract}

Information overload is a prevalent challenge in many high-value domains. A prominent case in point is the explosion of the biomedical literature on COVID-19, which swelled to hundreds of thousands of papers in a matter of months. In general, biomedical literature expands by two papers every minute, totalling over a million new papers every year. Search in the biomedical realm, and many other vertical domains is challenging due to the scarcity of direct supervision from click logs. 
Self-supervised learning has emerged as a promising direction to overcome the annotation bottleneck. 
We propose a general approach for vertical search based on domain-specific pretraining and present a case study for the biomedical domain. 
Despite being substantially simpler and not using any relevance labels for training or development, our method performs comparably or better than the best systems in the official TREC-COVID evaluation, a COVID-related biomedical search competition. 
Using distributed computing in modern cloud infrastructure, our system can scale to tens of millions of articles on PubMed and has been deployed as \DNAME, a new search experience for biomedical literature: \url{https://aka.ms/biomedsearch}.

\end{abstract}

%% file: 1_introduction.tex
\section{Introduction}

\eat{
Outline:
-	Motivation: Covid; complex search intent (maybe explicit or implicit reference to CC queries); challenges in vertical search (lack of query log for direct supervision)
-	Our approach
o	Model: neural language model + few-shot learning
	PubMedBERT, MetaAdaptRank, SaliencyMeasure
o	Index: PubMed, PMC, Covid
o	Scalability (L1/BERT ranker on Azure infrastructure) // good latency w/o distillation at low budget
o	UI
-	Evaluation: 
o	TREC submission
o	Ablation: compare impact of BERTs; few-shot; L1(?)
o	Curathon on CC queries(?) // highlight long query comparison
-	Discussion
o	General web search vs keyword-based (PubMed) vs vertical search
o	Scope / limitation / Future work
}

\begin{figure*}[t]
    \centering
    \includegraphics[width=0.8\linewidth]{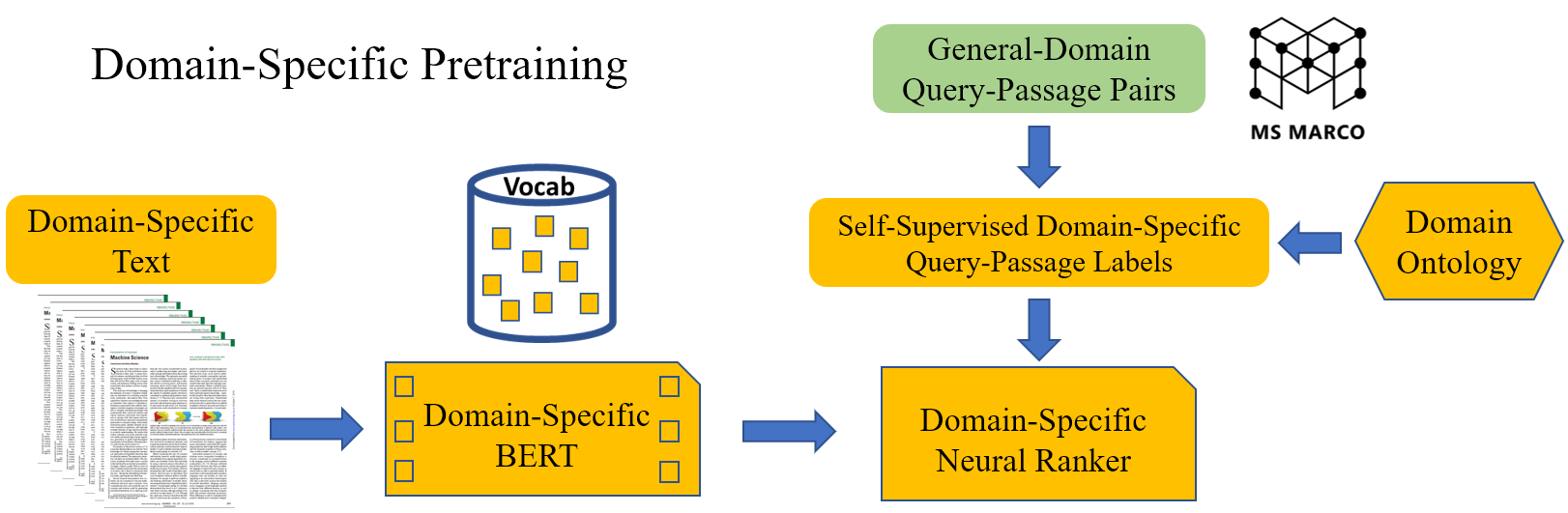} 
    \caption{
    General approach for vertical search: A neural ranker is initialized by domain-specific pretraining and fine-tuned on self-supervised relevance labels generated using a domain-specific lexicon from the domain ontology to filter query-passage pairs from MS MARCO.
    }
    \label{fig:vertical-search}
    \vspace{-3mm}
\end{figure*}

Keeping up with scientific developments on COVID-19 highlights the perennial problem of information overload in a high-stakes domain. At the time of writing, hundreds of thousands of research papers have been published concerning COVID-19 and the SARS-CoV-2 virus. For biomedicine more generally, the PubMed\footnote{\url{http://pubmed.ncbi.nlm.nih.gov}\label{fn:pubmed}} service adds 4,000 papers every day and over a million papers every year. 
While progress in general search has been made using sophisticated machine learning methods, such as neural retrieval models, vertical search is often limited to comparatively simple keyword search augmented by domain-specific ontologies (e.g., entity acronyms). The PubMed search engine exemplifies this experience. 
Direct supervision, while available for general search in the form of relevance labels from click logs, is typically scarce in specialized domains, especially for emerging areas such as COVID-related biomedical search. 

Self-supervised learning has emerged as a promising direction to overcome the annotation bottleneck, based on automatically creating noisy labeled data from unlabeled text. In particular, neural language model pretraining, such as BERT~\cite{devlin2018bert}, has demonstrated superb performance gains for general-domain information retrieval \cite{yang2019simple,nogueira2019passage, xiong2021approximate, lin2020pretrained} and natural language processing (NLP)~\cite{wang19iclr_glue,wang2019superglue}. Additionally, for specialized domains, domain-specific pretraining has proven to be effective for in-domain applications~\cite{lee2020biobert,beltagy2019scibert,pubmedbert,gururangan2020dapt,alsentzer-etal-2019-publicly,si&al19,huang2019clinicalbert}.

We propose a general methodology for developing vertical search systems for specialized domains. As a case study, we focus on biomedical search. We find evidence that the methods have significant impact in the target domain, and, likely generalize to other vertical search domains. We demonstrate how advances described in earlier and related work~\cite{pubmedbert,xiong2020cmt,zhang2020reinfoselect} can be brought together to provide new capabilities. We also provide data supporting the feasibility of a large-scale deployment through detailed system analysis, stress-testing of the system, and acquisition of expert relevance evaluations.\footnote{The system has been released, though large-scale deployment measures other than stress-testing are not yet available and we focus on the evidence from expert evaluation.}  

In \autoref{sec:self-sup}, we explore the key idea of initializing a neural ranking model with domain-specific pretraining and fine-tuning the model on a self-supervised domain-specific dataset generated from general query-document pairs (e.g., from MS MARCO~\cite{nyugen2016msmarco}). Then, we introduce the biomedical domain as a case study. 
%
In \autoref{sec:covid-search},
we evaluate the method on the TREC-COVID dataset~\cite{roberts2020trec,voorhees2020trec}. We find that the method performs comparably or better than the best systems in the official TREC-COVID evaluation, despite its generality and simplicity, and despite using zero COVID-related relevance labels for direct supervision. 
%
In \autoref{sec:pubmed-scale},
we discuss how our system design leverages distributed computing and modern cloud infrastructure for scalability and ease of use. This approach can be reused for other domains. In the biomedical domain, our system can scale to tens of millions of PubMed articles and attain a high query-per-second (QPS) throughput.
We have deployed the resulting system for preview as \DNAME, which provides a new search experience over biomedical literature: \url{https://aka.ms/biomedsearch}.


\eat{
A unique feature of COVID-19 Biomed Search is that it includes more biomedical research other than COVID-19 Open Research Dataset. In the multi-stage search architecture, we propose dense retrieval to solve vocabulary mismatch at retrieval stage. At ranking stage, given limited labeled data in COVID-19 domain, we propose continuous MLM training and data selection by weak supervision. These techniques have been verified in TREC-COVID challenge and achieve the highest fully automatic run in round 2. The website is publicly accessible via 
{\DNAME} is publicly available at \url{https://aka.ms/biomedsearch}

COVID-19 related scientific articles have been rapidly increasing as a response to the pandemics burst. The Allen Institute for AI (AI2) has released the COVID-19 Open Research Dataset (CORD-19), which contains 203,128 articles as of 7/29/2020. To help stakeholders to keep up with such fast-growing volume of COVID-19 related information, we present {\DNAME}, which is a search engine for COVID-19 and biomedical articles. 

There are some other COVID-19 search engines, for example, Covidex. These search engines rely on  CORD-19 dataset. However, during our collaboration with medical experts, we find the search need of users often go beyond the CORD-19. The search need often involves general biomedical knowledge, which indirectly links back to this specific pandemics . To accomplish such need, we collect a larger dataset of biomedical domain research literature, which are gathered from multiple data sources, including PubMed, PMC and preprints. To best of our knowledge, we are the first COVID-19 search engine to serve such dataset. The details of the dataset would be introduced later.

During the development of this system, we address several challenges ranging from research, evaluation and engineering. Below are our solutions:

1. Our system adopts standard multi-stage search architecture. At the retrieval stage, we apply keyword based retrieval method to generate the initial candidate paragraph pools for later stage. Although keyword based retrieval is lightweight, our enlarged data set still introduces additional challenge. The enlarged data set contains about 20 million articles, which is 100x bigger than CORD-19 dataset. In order to serve users at reasonable latency, we create a distributed retrieval layer using Elastic Search. We also propose another retrieval method, which encodes documents into dense vector via neural networks and use ANN to retrieve candidates at TREC-COVID challenge.

2. We leverage neural ranking models at ranking stage. Given limited labeled data in COVID-19 domain, we proposes continuous MLM training to train domain-specific language model using unlabeled data. Then we select similar domain data from web click data via weak supervision. 

3. We build user friendly interface on top of the backend algorithms and create {\DNAME}. The website is deployed at \hyperlink{http://biomedsearch.microsoft.com}. 

The overall workflow is described in Figure~\ref{fig:workflow}.

\begin{figure}[ht]
	\centering
 	\includegraphics[width=1.\textwidth]{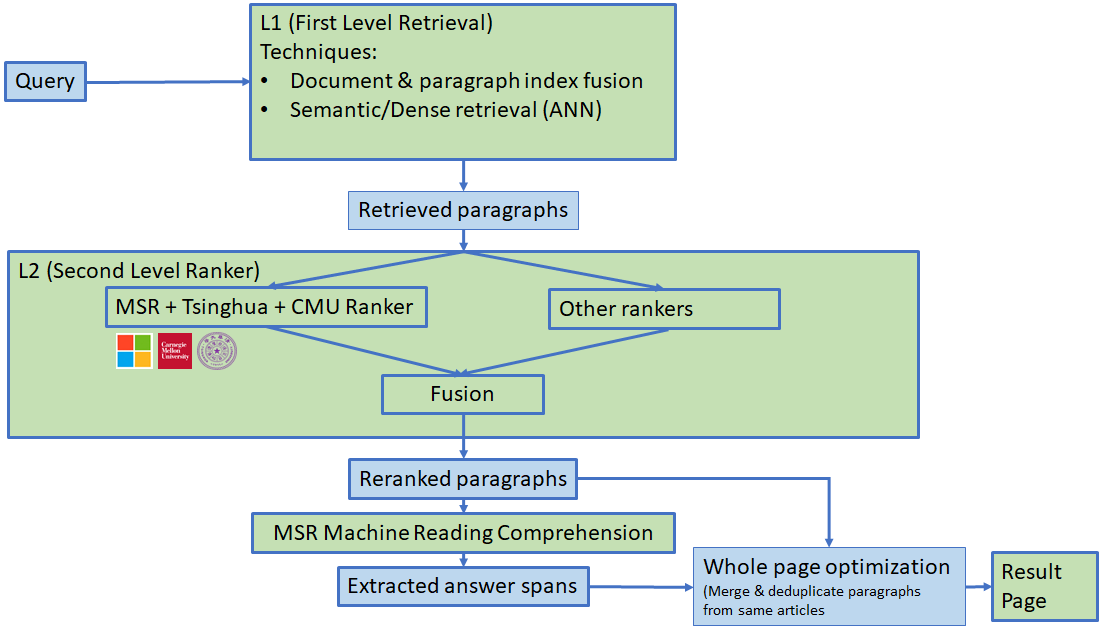}

\vspace{-0.2cm}
	\caption{The workflow of {\DNAME}: Retrieval stage (L1) retrieves candidate paragraphs with  paragraph indices. The main ranker in ranking stage (L2) is a neural ranking model. We also allow multiple rankers to work parallel and fuse their results to decide the final ranking order. Machine reading comprehesion component is applied to paragraphs to extract potential answers in case the query can be answered by a text span. Finally we group paragraphs by documents.}
	\label{fig:workflow}
\vspace{-0.2cm}	
\end{figure}

Also mention AWS COVID search (https://aws.amazon.com/blogs/publicsector/aws-launches-machine-learning-enabled-search-capabilities-covid-19-dataset/)

Also mention Google effort: https://blog.google/technology/health/making-data-useful-public-health/amp/?s=09

Chan-Zuckerberg (https://cziscience.medium.com/extracting-knowledge-from-biomedical-literature-402b4bed680a) and corresponding search tool: http://kdcovid.nl/

}

%% file: 2_self-sup.tex

\section{Domain-Specific Pretraining for Vertical Search}
\label{sec:self-sup}

In this section, we present a general approach for vertical search based on domain-specific pretraining and self-supervised learning (\autoref{fig:vertical-search}). 
We first review neural language models and show how domain-specific pretraining can serve as the foundation for a domain-specific document neural ranker. We then present a general method of fine-tuning the ranker by using self-supervised, domain-specific relevance labels from a broad-coverage query-document dataset using the domain ontology. Finally, we show how this approach can be applied in biomedical literature search.

\subsection{Domain-Specific Pretraining}

Language model pretraining can be considered a form of \textit{task-agnostic self-supervision} that generates training examples by hiding words from unlabeled text and tasks the model with predicting the hidden words. In our work on vertical search, we adopt the popular Bidirectional Encoder Representations from Transformers~(BERT) \cite{devlin2018bert}, which has become a standard building block for NLP applications.  
Instead of predicting the next token based on the preceding tokens, as in traditional generative models, BERT employs a \textit{Masked Language Model~(MLM)}, which randomly replaces a subset of tokens by a special token $[MASK]$, and tries to predict them from the rest of the words.  The training objective is the cross-entropy loss between the original tokens and the predicted ones. 
BERT builds on the transformer model~\cite{vaswani2017attention} with its multi-head self-attention mechanism, which has demonstrated high performance in parallel computation and modeling long-range dependencies, as compared to recurrent neural networks such as LSTM~\cite{hochreiter1997lstm}. 
The input consists of text spans, such as sentences, separated by a special token~$[SEP]$. To address out-of-vocabulary words, tokens are divided into subword units using Byte-Pair Encoding~(BPE) \cite{sennrich2015bpe} or its variants \cite{kudo2018sentencepiece}, which generates a fixed-size subword vocabulary to compactly represent the training text corpora. 
The input is first passed to a lexical encoder, which combines the token embedding, position embedding, and segment embedding by element-wise summation. 
The embedding layer is then passed to multiple layers of transformer modules to generate a contextual representation~\cite{vaswani2017attention}. 

Prior pretraining efforts have focused frequently on the newswire and web domains. For example, the BERT model was trained on Wikipedia\footnote{\url{http://wikipedia.org}} and BookCorpus~\cite{zhu&al15_books}, and subsequent efforts have focused on crawling additional web text to conduct increasingly large-scale pretraining~\cite{liu2019roberta,raffel2019t5,brown2020gpt3}.
For domain-specific applications, pretraining on in-domain text has been shown to provide additional gains, but the prevalent assumption is that out-domain text is still helpful and pretraining typically adopts a mixed-domain approach~\cite{lee2020biobert,gururangan2020dapt}. \citet{pubmedbert} changes this assumption and shows that, for domains with ample text, a pure domain-specific pretraining approach is advantageous and leads to substantial gains in downstream in-domain applications. We adopt this approach by generating domain-specific vocabulary and performing language model pretraining from scratch on in-domain text~\cite{pubmedbert}.

\subsection{Self-Supervised Fine-Tuning}

As a first-order approximation, the search problem can be abstracted as learning a relevance function for query $q$ and text span $t$: $f(q,t)\rightarrow \{0,1\}$. Here, $t$ may refer to a document or arbitrary text span such as a passage. 

Traditional search methods adopt a sparse retrieval approach by essentially treating the query as a bag of words and matching each word against the candidate text, which can be done efficiently using an inverted index. Individual words are weighted (e.g., by TF-IDF) to downweight the effect of stop words or function words, as exemplified by BM25 and its variants \cite{robertson1976bm25}. 

Variations abound in natural language expressions, which can cause significant challenges in sparse retrieval. To address this problem, dense retrieval maps query and text each to a vector in a continuous representation space and estimates relevance by computing the similarity between the two vectors (e.g., via dot product)~\cite{karpukhin2020dense,huang2013learning,xiong2021approximate}.
Dense retrieval can be made highly scalable by pre-computing text vectors, and can potentially replace or combine with sparse retrieval.

Neither sparse retrieval nor dense retrieval attempts to model complex interdependencies between the query and text. 
In contrast, sophisticated neural approaches concatenate query and text as input for a BERT model to leverage cross-attention among query and text tokens \cite{yilmaz2019applying}. 
Specifically, query $q$ and text $t$ are combined into a sequence ``$[CLS]$ q $[SEP]$ t $[SEP]$'' as input, where $[CLS]$ is a special token to be used for final prediction \cite{devlin2018bert}. 
This could produce significant performance gains but requires a large amount of labeled data for fine-tuning the BERT model. 
Such a cross-attention neural model will not be scalable enough for the retrieval step, as we must compute, from scratch, for each candidate text with a new query. The standard practice thus adopts a two-stage approach, by using a fast L1 retrieval method to select top $K$ text candidates, and applying the neural ranker on these candidates as L2 reranking. 

In our proposed approach, we use BM25 for L1 retrieval, and initialize our L2 neural ranker with a domain-specific BERT model. 
To fine-tune the neural ranker, we use the Microsoft Machine Reading Comprehension dataset, MS MARCO~\cite{nyugen2016msmarco}, and a domain-specific lexicon to generate noisy relevance labels at scale using self-supervision (\autoref{fig:vertical-search}). MS MARCO was created by identifying pairs of anonymized queries and relevant passages from Bing's search query logs, and crowd-sourcing potential answers from passages. 
The dataset contains about one million questions spanning a wide range of topics, each with corresponding relevant answer passages from Bing question answering systems.
For self-supervised fine-tuning labels, we use the MS MARCO subset~\cite{macavaney2020sledge} whose queries contain at least one domain-specific term from the domain ontology.

\vspace{-3mm}
\subsection{Application to Biomedical Literature Search}
\label{sec:biomed-search}

\eat{
\begin{figure}
    \centering
    \includegraphics[width=\linewidth]{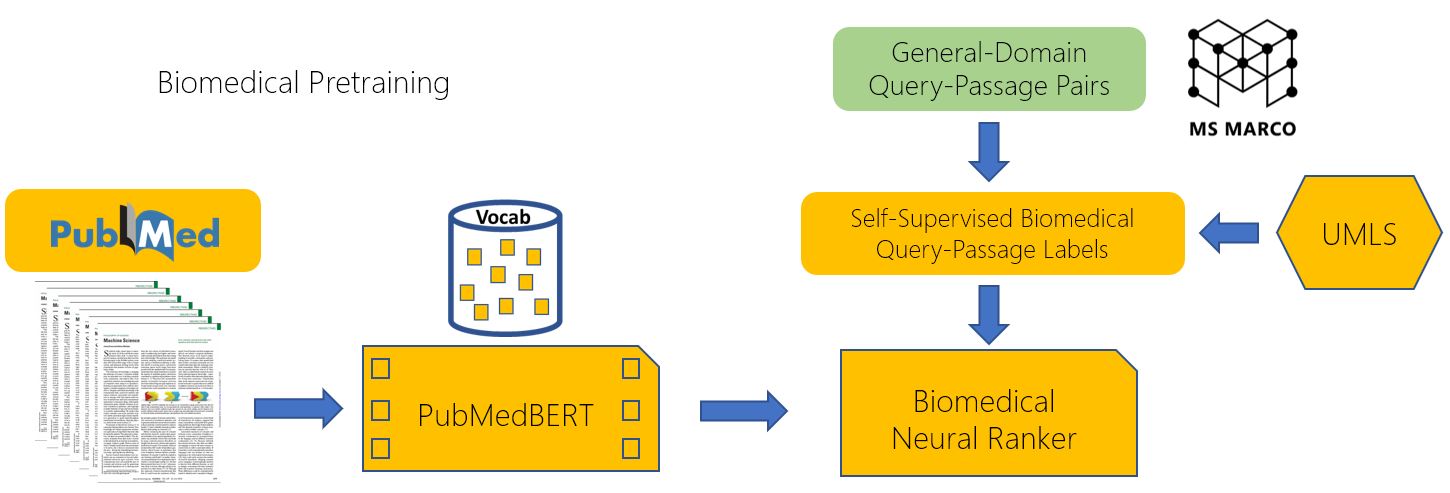}
    \caption{
    Example application: Biomedical search. Domain-specific pretraining is conducted on a corpus of over 30 million PubMed articles. Self-supervised relevance labels are generated from MS MARCO queries that contain disease or syndrome terms in the Unified Medical Language System (UMLS).
    }
    \label{fig:biomed-search}
\end{figure}
}

Biomedicine is a representative case study that illustrates the challenges of vertical search. It is a high-value domain with a vast and rapidly growing research literature, as evident in PubMed (30+ million articles; adding over a million a year). However, existing biomedical search tools are typically limited to sparse retrieval methods, as exemplified by PubMed. This search is primarily limited to keyword matching, though it is augmented with limited query expansion using domain ontologies (e.g., MeSH terms~\cite{lipscomb2000mesh}). This method is suboptimal for long queries expressing complex intent. 

We use biomedicine as a running example to illustrate our approach for vertical search. 
We leverage PubMed articles for domain-specific pretraining and use the publicly-available PubMedBERT~\cite{pubmedbert} to initialize our L2 neural ranker. For self-supervised fine-tuning, we use the Unified Medical Language System~(UMLS)~\cite{bodenreider2004umls} as our domain ontology and filter MS MARCO queries using the disease or syndrome terms in UMLS, similar to \citet{macavaney2020sledge,macavaney-etal-2020-sledgez} but focusing on the broad biomedical literature rather than COVID-19. 
This medical subset of MS MARCO contains about 78 thousand annotated queries. 
We used these queries and their relevant passages in MS MARCO as positive relevance labels.
To generate negative labels, we ran BM25 for each query over all non-relevant passages in MS MARCO, and selected the top 100 results. This forces the neural ranker to work harder in separating truly relevant passages from ones with mere overlap in keywords. For balanced training, we down-sampled negative instances to equal the number of positive instances (i.e., 1:1 ratio). This resulted in about 640 thousand (query, passage, label) examples.

Based on preliminary experiments, we chose a learning rate of $2e-5$ and ran fine-tuning for one epoch in all subsequent experiments. We found that the results are not sensitive to hyperparameters, as long as the learning rate is of the same order of magnitude and at least one epoch is run over all the examples. 
At retrieval time, we used $K=60$ in the L1 ranker by default (i.e., we used BM25 to select top 60 text candidates).


\eat{
medical MARCO:
To produce this subset, we use the MedSyn
lexicon (Yates and Goharian, 2013), which includes
layperson terminology for various medical conditions. Only queries that contain terms from the lexicon are considered in this dataset, leaving 78,895
of the original 808,531 training queries (9.7\%).5 
}

%% file: 3_covid.tex
\section{Case Study Evaluation on COVID-19 Search}
\label{sec:covid-search}

The COVID-19 literature provides a realistic test ground for biomedical search. In a little over a year, the COVID-related biomedical literature has grown to include over 440 thousand papers that  mention COVID-19 or the SARS-CoV-2 virus. This explosive growth sparked the creation of the COVID-19 Open Research Dataset (CORD-19)\cite{wang2020cord19} and subsequently TREC-COVID~\cite{roberts2020trec,voorhees2020trec}, an evaluation resource for pandemic information retrieval. 

In this section, we describe our evaluation of the biomedical search system on TREC-COVID, focusing on two key questions. First, how does our system perform compared to the best systems participating in TREC-COVID? We note that many of these systems are expected to have complex designs and/or require COVID-related relevance labels for training and development. 
Second, what is the impact of domain-specific pretraining compared to general-domain or mixed-domain pretraining?


\subsection{The TREC-COVID Dataset}

To create TREC-COVID, organizers from the National Institute of Standards and Technology (NIST) used versions of CORD-19 from April 10 (Round 1), May 1 (Round 2), May 19 (Round 3), June 19 (Round 4), and July 16 (Round 5).
These datasets spanned an initial set of 30 topics with five new topics planned for each additional round; the final set thus consists of 50 topics and cumulative judgements from previous rounds generated by domain experts~\cite{roberts2020trec}.
Relevance labels were created by annotators using a customized platform and released in rounds. Round 1 contains 8,691 relevance labels for 30 topics, and was provided to participating teams for training and development. 
Subsequent rounds were hosted to introduce additional topics and relevance labels as a rolling evaluation for increased participation.
We use Round 2, the round we participated in, to evaluate our system development.
It contains 12,037 relevance labels for 35 topics.
 

\subsection{Top Systems in TREC-COVID Leaderboard}

The results of TREC-COVID Round 2 are organized into three groups: {\em Manual}, which used manual interventions, e.g., manual query rewriting, in any part of the system, {\em Feedback}, which used labels from Round 1, and {\em Automatic}, which does not use manual effort or Round 1 labels.\footnote{\url{https://castorini.github.io/TREC-COVID/round2/}} Note that the categorization of {\em Feedback} and {\em Automatic} is not always explicit so their grouping might be mixed. Overall, 136 systems participated in the official evaluation. 
NDCG@10 was used as the main evaluation metric, with Precision@5 (P@5) reported as an additional metric. 
The best performing systems typically adopted a sophisticated neural ranking pipeline and performed extensive training and development on TREC-COVID labeled data from Round 1. 
Some systems also use very large pretrained language models. 
For example, $\tt covidex.t5$ used T5 Large~\cite{raffel2019t5}, a general-domain transformer-based model with 770 million parameters pretrained on the Colossal Clean Crawled~(C4) web corpus (26 TB).\footnote{\url{https://www.tensorflow.org/datasets/catalog/c4}}

The best performing non-manual system for Round 2 is CMT (CMU-Microsoft-Tsinghua)~\cite{xiong2020cmt}, which adopted a two-stage ranking approach. 
For L1, CMT used standard BM25 sparse retrieval as well as dense retrieval by fusing top ranking results from the two methods. The dense retrieval method computed the dot product of query and passage embeddings based on a BERT model~\cite{karpukhin2020dense}. 
For L2, CMT used a neural ranker with cross-attention over query and candidate passage. 

For training, CMT started with the same biomedical MS MARCO data (by selecting MS MARCO queries with biomedical terms)~\cite{macavaney2020sledge}, but then applied additional processing to generate synthetic labeled data. Briefly, it first trained a query generation system using \emph{query generation} (QG)~\cite{nogueira2019passage} on the query-passage pairs from biomedical MS MARCO, initialized by GPT-2~\cite{radford2019gpt2}. Given this trained QG system, for each COVID-related document $d$, it generated a pseudo query $q=QG(d)$, and then applied BM25 to retrieve a pair of documents with high and low ranking, $d'_+, d'_-$. Finally, it called on ContrastQG~\cite{xiong2020cmt} to generate a query that would best differentiate the two documents $q'=ContrastQG(d'_+,d'_-)$. 
For the neural ranker, CMT started with SciBERT~\cite{beltagy2019scibert} with continual pretraining on CORD-19, and fine-tuned the model using both Med MARCO labels and synthetic labels from ContrastQG. 

To leverage the TREC-COVID data from Round 1, CMT incorporated data reweighting (ReinfoSelect) based on the REINFORCE algorithm \cite{zhang2020reinfoselect}. It used performance on Round 1 data as a reward signal, and learned to denoise training labels by re-weighting them using policy gradient. 

\subsection{Our Approach on TREC-COVID}

\begin{table}[t!]
    \centering
    \setlength{\tabcolsep}{1.0mm}{
    \begin{tabular}{lcccc}
    \toprule
Model & NDCG@10 & P@5   \\
\midrule
\multicolumn{3}{l}{{\hspace{-1mm} \em Our approach:  }}  \\
PubMedBERT& 61.5 ($\pm$1.1) & 69.5 ($\pm$1.8) \\
PubMedBERT-COVID& 65.6 ($\pm$1.0) & 73.2 ($\pm$1.1)\\
\midrule
\multicolumn{3}{l}{{\hspace{-1mm} \em + dev set: }}  \\
PubMedBERT & 64.8 &  71.4 \\
PubMedBERT-COVID & 67.9 &  73.7 \\
\midrule
\multicolumn{3}{l}{{\hspace{-1mm} \em Top systems in TREC-COVID: }}  \\
covidex.t5 (T5) & 62.5 & 73.1 \\
mpiid5 (ELECTRA) & 66.8 &  77.7 \\
CMT (SparseDenseSciBERT) & 67.7 &  76.0 \\
    \bottomrule
    \end{tabular}
    }
    \caption{Comparison with the top-ranked systems in official TREC-COVID evaluation (test results; Round 2). Our results were averaged from ten runs with different random seeds (standard deviation shown in parentheses). The best systems in TREC-COVID evaluation (bottom panel) all used Round 1 data for training, as well as more sophisticated learning methods and/or larger models such as T5. 
    In contrast, our systems (top panel) are much simpler and used zero TREC-COVID relevance labels, but they already perform competitively against the best systems by using domain-specific pretraining (PubMedBERT). 
    Our systems were trained using one epoch with a fixed learning rate. By exploring longer training and multiple learning rates and using Round 1 data for development, our systems can perform even better (middle panel).}
    \label{tbl:compare-top-sys}
    \vspace{-5mm}
\end{table}

\begin{table}[t!]
    \centering
    \setlength{\tabcolsep}{1.0mm}{
    \begin{tabular}{lcccc}
    \toprule
Model & NDCG@10 & P@5   \\
\midrule
BERT & 55.0 ($\pm$1.2) & 63.4 ($\pm$2.3)\\
RoBERTa & 53.5 ($\pm$1.6) & 61.1 ($\pm$2.3)\\
UNILM & 55.0 ($\pm$1.2) & 62.0 ($\pm$1.8)\\
SciBERT & 58.9 ($\pm$1.5) & 67.7 ($\pm$2.2)\\
PubMedBERT& 61.5 ($\pm$1.1) & 69.5 ($\pm$1.8) \\
PubMedBERT-COVID& {\bf 65.6} ($\pm$1.0) & {\bf 73.2} ($\pm$1.1)\\
    \bottomrule
    \end{tabular}
    }
    \caption{Comparison of domain-specific (PubMedBERT and PubMedBERT-COVID) pretraining with out-domain (BERT, RoBERTa, UniLM) or mixed-domain pretraining (SciBERT) in TREC-COVID test results (Round 2). All results were averaged from ten runs (standard deviation in parentheses). Domain-specific pretraining is essential for attaining good performance in our general approach for vertical search.
    }
    \label{tbl:bert-compare}
    \vspace{-5mm}
\end{table}

\begin{table*}[]
\begin{tabular}{l|ccc|l|l}
\hline
{\bf Biomedical Search Engine} & {\bf CORD-19} & {\bf PubMed} & {\bf PMC} & {\bf Retrieval} & {\bf Reranking} \\ \hline
PubMed\footnote{\url{https://pubmed.ncbi.nlm.nih.gov/}} & & \checkmark &  & Keyword + MeSH & \\ \hline
COVID-19 Search (Azure)\footnote{\url{https://covid19search.azurewebsites.net/home/index?q=}} & \checkmark &  &  & BM25 & \\ \hline
CORD-19 Explorer (AI2)\footnote{\url{https://cord-19.apps.allenai.org/}} & \checkmark & & & BM25 & LightGBM \\ \hline
COVID-19 Research Explorer (Google)\footnote{\url{https://covid19-research-explorer.appspot.com/}} & \checkmark  & &  & BM25 & Neural (BERT) \\ \hline
Covidex (U of Waterloo, NYU)\footnote{\url{https://covidex.ai/}} & \checkmark &  &  & BM25 & Neural (T5)\\ \hline
COVID-19 Search (Salesforce)\footnote{\url{https://sfr-med.com/search}} & \checkmark & & & BM25 & Neural (BERT) \\ \hline
Microsoft Biomedical Search\footnote{\url{https://https://aka.ms/biomedsearch/}} & \checkmark &\checkmark &\checkmark & BM25 & Neural (PubMedBERT) \\
\bottomrule
\end{tabular}
\caption{Overview of representative biomedical search systems. $\checkmark$ signifies coverage on CORD-19 (440 thousand abstracts and full-text articles), PubMed (30 million abstracts), PubMed Central (PMC; 3 million full-text articles). Most systems cover CORD-19 (or the earlier version with about 60 thousand articles). 
Only Microsoft Biomedical Search (our system) uses domain-specific pretraining (PubMedBERT), which outperforms general-domain language models, for neural reranking.}
    \label{tbl:biomed-search-systems}
    \vspace{-5mm}
\end{table*}


TREC-COVID offers an excellent benchmark for assessing the general applicability of our proposed approach for vertical search. We evaluated our systems on the test set (Round 2) and compared them with the best systems in the official TREC-COVID evaluation. We essentially took the biomedical search system from \autoref{sec:biomed-search} as is (PubMedBERT). Although COVID-related text may differ somewhat from general biomedical text, we expect that a biomedical model should offer strong performance for this subset of biomedical literature.
To further assess the impact from domain-specific pretraining, we also conducted continual pretraining using CORD-19 for 100K BERT steps and evaluated it in our biomedical search system (PubMedBERT-COVID).

\autoref{tbl:compare-top-sys} shows the results. Surprisingly, without using any relevance labels, our systems (top panel) performs competitively against the best systems in TREC-COVID evaluation. E.g., PubMedBERT-COVID outperforms $\tt covidex.t5$ by over three absolute points in NDCG@10, even though the latter used a much larger language model 
pretrained on three orders of magnitude more data (26TB vs 21GB). 
Our systems were trained using one epoch with a fixed learning rate (2e-5). By exploring longer training (up to five epochs) and multiple learning rates (1e-5, 2e-5, 5e-5) and using Round 1 as dev set, our best system (middle panel) performs on par in NDCG@10 with CMT, the top system in TREC-COVID, while requiring no additional sophisticated learning components such as dense retrieval, QG, ContrastQG, and ReinfoSelect. 

The success of our systems can be attributed primarily to our in-domain language models (PubMedBERT, PubMedBERT-COVID). To further assess the impact of domain-specific pretraining, we also evaluated our system using out-domain and mixed-domain models. See \autoref{tbl:bert-compare} for the results. Out-domain language models all perform relatively poorly in this evaluation of biomedical search, and exhibit little difference in search relevance despite significant difference in the size of vocabulary, pretraining corpus, and model (e.g., RoBERTa \cite{liu2019roberta} used a larger vocabulary and both RoBERTa and UniLM \cite{dong2019unified} were pretrained on much larger text corpus). 
Pretraining on PubMed text helps SciBERT, but its mixed-domain approach (including compute science literature) inhibits its performance compared to domain-specific pretraining. 
Continual pretraining on covid-specific literature helps substantially, with PubMedBERT-COVID outperforming PubMedBERT by over four absolute points in NDCG@10. 
Overall, domain-specific pretraining is essential for the performance gain, with PubMedBERT-COVID outperforming general-domain BERT models by over ten absolute points in NDCG@10. 

In sum, the TREC-COVID results provide strong evidence that, by leveraging domain-specific pretraining, our approach for vertical search is general and can attain high accuracy in a new domain without significant manual effort.

\eat{
\begin{table}[t!]
    \centering
    
    \setlength{\tabcolsep}{1.0mm}{
    \begin{tabular}{lcccc}
    \toprule
Model & NDCG@10 & P@5   \\
\midrule
\multicolumn{3}{l}{{\hspace{-1mm} \em round2 : }}  \\
SparseDenseSciBERT & 67.7 &  76.0 \\
mpiid5 (ELECTRA) & 66.8 &  77.7 \\
ReInfoSelect & 62.6 & 69.7 \\
covidex.t5 (T5) & 62.5 & 73.1 \\
\midrule
\multicolumn{3}{l}{{\hspace{-1mm} \em with dev set : }}  \\
BERT & 57.4  &  67.4\\
Roberta & 56.8 &  65.1\\
UNILM & 57.9 & 66.9\\
SciBERT &  63.9 &  74.3 \\
PubmedBERT & 64.8 &  71.4 \\
PubmedBERT-COVID & 67.9 &  73.7 \\
\midrule
\multicolumn{3}{l}{{\hspace{-1mm} \em without dev set (trained with 1 epoch):  }}  \\
BERT & 55.0 $\pm$1.2 & 63.4 $\pm$2.3\\
Roberta & 53.5 $\pm$1.6 & 61.1 $\pm$2.3\\
UNILM & 55.0 $\pm$1.2 & 62.0 $\pm$1.8\\
SciBERT & 58.9 $\pm$1.5 & 67.7 $\pm$2.2\\
PubmedBERT& 61.5 $\pm$1.1 & 69.5 $\pm$1.8 \\
PubmedBERT-COVID& 65.6 $\pm$1.0 & 73.2 $\pm$1.1\\
    \bottomrule
    \end{tabular}
    }
    \caption{}
    \label{}
    \vspace{-5mm}
\end{table}
}

    

%% file: 4_ms-biomed-search.tex


\begin{figure*}
    \centering
    \includegraphics[width=0.8\linewidth]{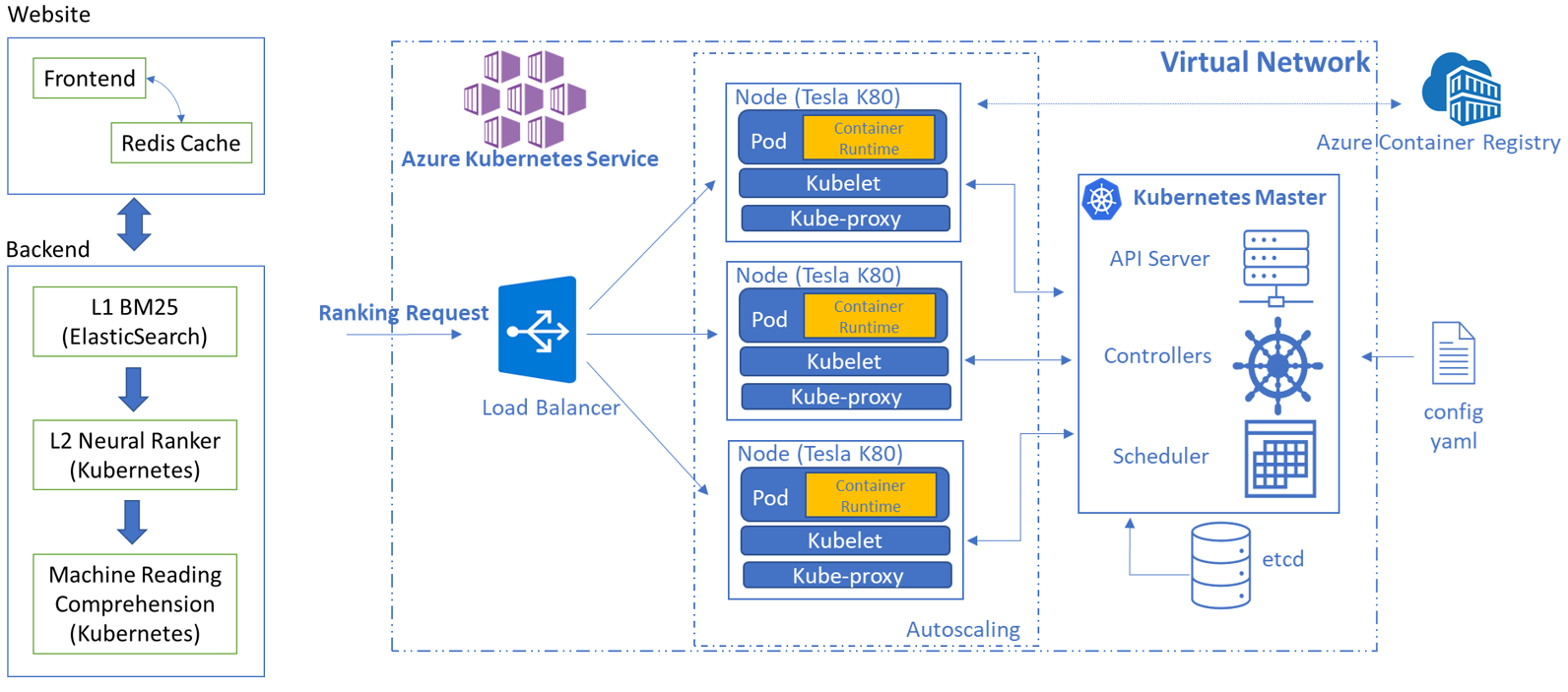}
    \caption{
    Left: Overview of the Microsoft Biomedical Search system. Right: A reference cloud architecture for servicing the L2 neural ranker and machine reading comprehension (MRC) with automatic scaling. Queries are processed by a standard two-stage architecture, where an L1 ranker based on BM25 generates the top 60 passages for each query, followed by an L2 neural ranker to produce final reranking results, which are then passed to the MRC module to generate answers from a candidate passage if applicable.
    }
    \label{fig:system}
    \vspace{-2mm}
\end{figure*}


\addtocounter{footnote}{-6}
\footnotetext{\url{https://pubmed.ncbi.nlm.nih.gov/}}
\addtocounter{footnote}{1}
\footnotetext{\url{https://covid19search.azurewebsites.net/home/index?q=}}
\addtocounter{footnote}{1}
\footnotetext{\url{https://cord-19.apps.allenai.org/}}
\addtocounter{footnote}{1}
\footnotetext{\url{https://covid19-research-explorer.appspot.com/}}
\addtocounter{footnote}{1}
\footnotetext{\url{https://covidex.ai/}}
\addtocounter{footnote}{1}
\footnotetext{\url{https://sfr-med.com/search}}
\addtocounter{footnote}{1}
\footnotetext{\url{https://aka.ms/biomedsearch}}

\section{PubMed-Scale Biomedical Search}
\label{sec:pubmed-scale}

The canonical tool for biomedical search is the PubMed search itself.
Recently, COVID-19 has spawned a plethora of new prototype biomedical search tools. See \autoref{tbl:biomed-search-systems} for a list of representative systems. 
PubMed covers essentially the entire biomedical literature, but its aforementioned search engine is based on relatively simplistic sparse retrieval methods, which generally perform less well, especially in the presence of long queries with complex intent. 
By contrast, while some new search tools feature advanced neural ranking methods, their search scope was typically limited to CORD-19, which considers only a tiny fraction of biomedical literature. 
In this section, we describe our effort in developing and deploying \DNAME, a new biomedical search engine that combines PubMed-scale coverage and state-of-the-art neural ranking, based on our general approach for vertical search, as described in \autoref{sec:biomed-search} and validated in \autoref{sec:covid-search}. Creating the system required addressing significant challenges with system design and engineering. Employing a modern cloud infrastructure helped with the fielding of the system. The fielded system can serve as a reference architecture for vertical search in general; many components are directly reusable for other high-value domains.

\subsection{System Challenges}

The key challenge in the system design is to scale to tens of millions of biomedical articles, while enabling affordable and fast computation in sophisticated neural ranking methods, based on large language models with hundreds of millions of parameters.

Specifically, the CORD-19 dataset initially covered about 29,000 documents (abstracts or full-text articles) when it was first launched in March 2020. It quickly grew to about 60,000 documents when it was adopted by TREC-COVID (Round 2, May 2020), which is the version used by many COVID-search tools. Even in its latest version (as of early Feb. 2021), CORD-19 only contains about 440,000 documents (with about 150,000 full-text articles). By contrast, PubMed covers over 30 million biomedical publications, with about 20 million abstracts and over 3 million full-text articles, which is two orders of magnitude larger than CORD-19. 

Given early feedback from a range of biomedical practitioners, in addition to document-level retrieval, we decided to enable passage-level retrieval to enhance granularity and precision. This further exacerbates our scalability challenge, as the retrieval candidates now include over 216 million paragraphs (passages).

Neural ranking methods can greatly improve search relevance compared to standard keyword-based and sparse retrieval methods. However, they present additional challenges as these methods often build upon large pretrained language models, which are computational intensive and generally require expensive graphic processing units (GPUs).

\vspace{-3mm}
\subsection{Our Solution}

As described in \autoref{sec:biomed-search}, we adopt a two-stage ranking model, with an L1 ranker based on BM25 and an L2 reranker based on PubMedBERT. As shown in \autoref{fig:system} (left), the system comprises a web front end, web back end API, cache, L1 ranking, and L2 ranking. 
Query requests are passed on from web front end to back end API, which coordinates L1 and L2 ranking. The system first consults the cache and returns results directly if the query is cached. Otherwise, it calls on L1 to retrieve top candidates and then calls on L2 to conduct neural reranking. Finally, it combines the results and returns them to the front end for display. 

To address the scalability challenges, we develop our system on top of modern cloud infrastructures to leverage their native capabilities of distributed computing, cache, and load balancing, which drastically simplifies our system design and engineering. We choose to use Microsoft Azure as the cloud infrastructure, but our design is general and can be easily adapted to other cloud infrastructures. 

In early experiments, we found that the Web front end, back end and cache components are sufficiently fast. So, in what follows, we will focus on discussing how to address scalability challenges in L1 and L2 ranking.

For L1, we use BM25, which can be supported by standard inverted index methods. We adopt Elastic Search, an open-source distributed search engine built on Apache Lucene~\cite{gormley2015elasticsearch}. 
Given our PubMed-scale coverage, the index size of Elastic Search is over 160GB and is growing as new papers arrive. The index size further multiplies with the number of replications added to ensure system availability (we use two replications). 
As such, we need to use machines with enough memory and processing power.

For L2, although we only run on limited number of candidate passages from L1 (we used top 60 in our system), the neural ranking model is based on large pretrained language models which are computationally intensive. Currently, we use the base model of PubMedBERT with 12 layers of transformer modules, containing over 300 million parameters. We thus use a distributed GPU cluster and make careful hardware and software choices to maximize overall cost-effectiveness while minimizing L2 latency.

We use query-per-second (QPS) as our key workload metric for system design. 
To identify major bottlenecks and fine-tune design choices, we conducted focused experiments on L1 and L2 rankers separately to assess their impact on run-time latency.  

\eat{
\begin{figure}
    \centering
    \includegraphics[width=\linewidth]{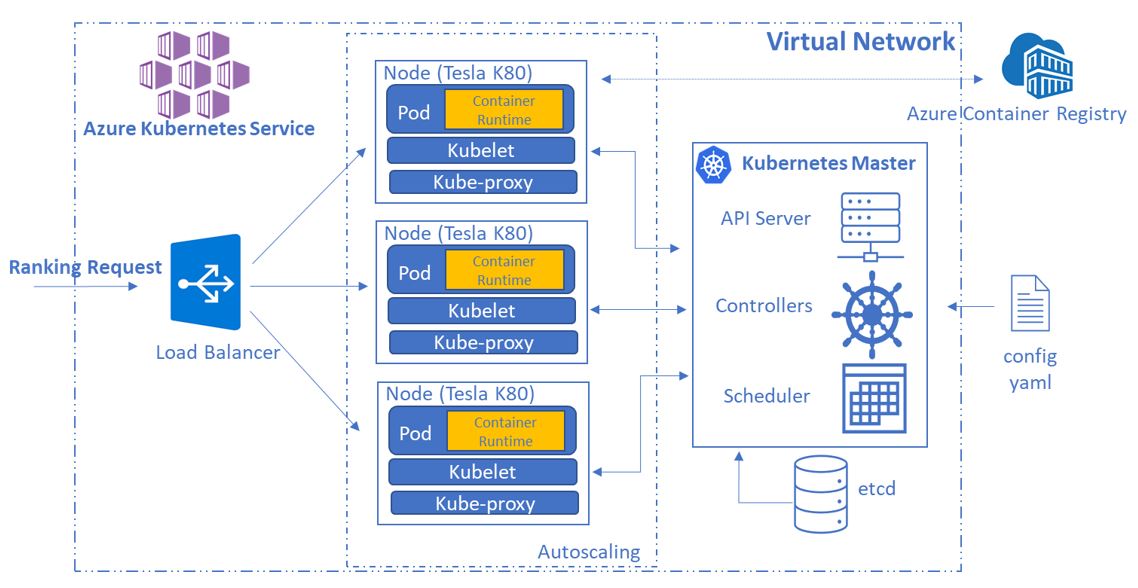}
    \caption{
    Reference cloud architecture for servicing the L2 neural ranker with automatic scaling.
    }
    \label{fig:L2-sys}
    \vspace{-3mm}
\end{figure}
}

We use Locust~\cite{locust}, a Python-based framework for load testing. To ensure head-to-head comparison among design choices, we adopted a fixed system setup as follows:
\begin{itemize}
    \item The back end API is developed with Flask~\cite{flask}, using Gevent~\cite{gevent} with 8 workers to ensure the highest performance
    \item To minimize variance due to network cost, the back end API and L1 or L2 rankers are deployed in the same data center, as well as machines used to send queries.
    \item All the servers are deployed in the same virtual network.
    \item We prepare a query set which contains 71 thousand anonymized queries sampled from Microsoft Academic Search. 
    \item We turned off the cache layer during all experiments.
\end{itemize}

With this configuration, the latency of back end API per query is around 20 ms. We used the Locust client to simulate asynchronous requests from multiple users. Each simulated user would randomly wait for 15-60 seconds after each search request. Each experiments ran for 10 minutes. 

From preliminary experiments, we found that Elastic Search requires warm-up to reach maximum performance, so we ran the system with low QPS (0.5 per sec) for 10 minutes before conducting the our experiments. Elastic Search might cache results to speed up repeat queries. 
To eliminate confounders from caching, we ensure that no query is repeated in each experiment.

For L1, based on the performance experiments, we chose the following configuration for Elastic Search:
\begin{itemize}
    \item Each query is processed by a main node, which distributes its query to data nodes and then merges the results. 
    \item There are three main nodes and ten data nodes, each using a premium machine (D8s v3) with a 1TB SSD disk (P30). 
    \item The index is divided into 30 shards.
\end{itemize}

\begin{figure*}
    \centering
    \includegraphics[width=0.85\linewidth]{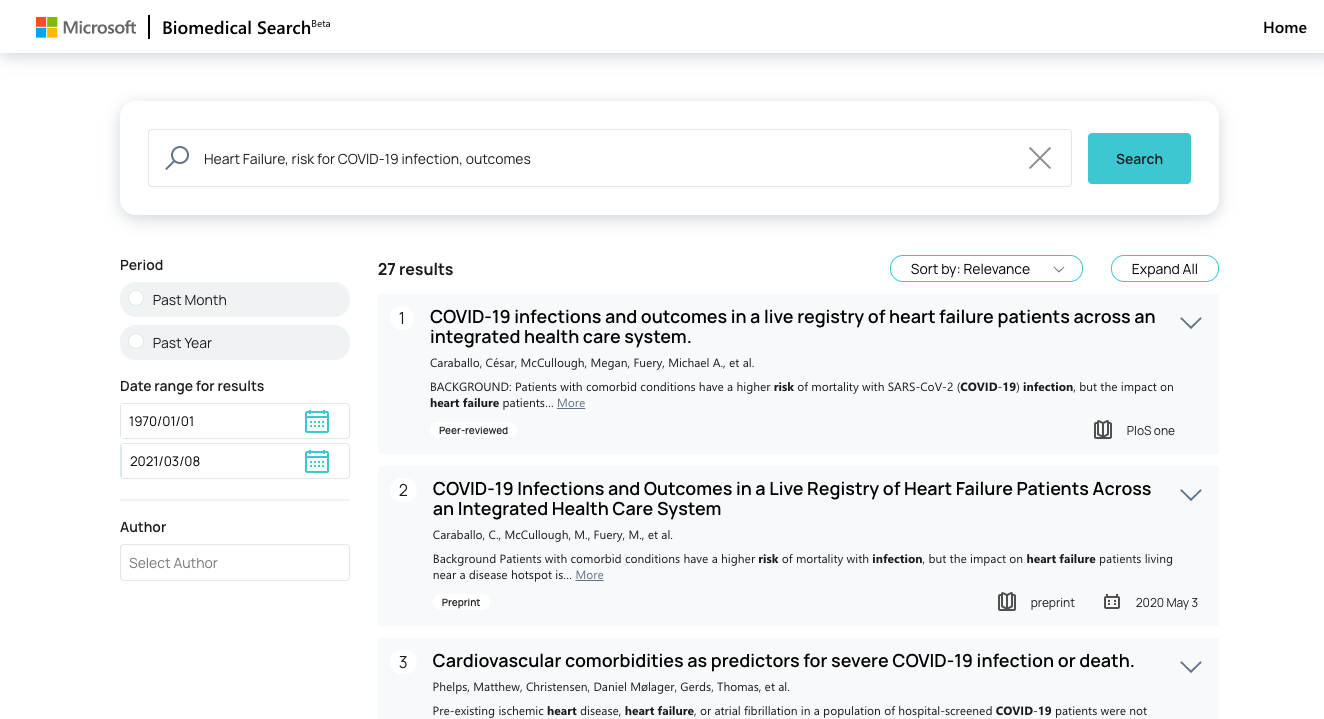}
    \caption{Sample screenshot of Microsoft Biomedical Search. The system applies our general approach for vertical search based on domain-specific pretraining and self-supervision, and covers all abstracts and full-text articles in CORD-19, PubMed, and PubMed Central (PMC).}
    \vspace*{-\baselineskip}
    \label{fig:ux}
\end{figure*}

\begin{table}[t!]
    \centering
    \setlength{\tabcolsep}{1.0mm}{
    \begin{tabular}{cccccc}
    \toprule
QPS & Median (s) & 90\% (s) & Mean (s) & Min (s) & Max (s)  \\
\midrule
13.2 &	0.51	& 0.75 &	0.59 & 0.23 &	7.07\\
26.8 &	0.60 & 1.50 & 0.88 & 0.22 & 31.0 \\
    \bottomrule
    \end{tabular}
    }
    \caption{Latency results in two simulated load tests on L1 ranking (plus back end API). Query-per-second (QPS) is the average request load in the test. Back end API takes about 20 ms for each query. Most queries can be processed within a second, even with relatively high request load.}
    \label{tbl:L1-test}
    \vspace{-5mm}
\end{table}

\begin{table}[t!]
    \centering
    \setlength{\tabcolsep}{1.0mm}{
    \begin{tabular}{cccccc}
    \toprule
QPS & Median (s) & 90\% (s) & Mean (s) & Min (s) & Max (s)  \\
\midrule
14.8 &	1.80 &	2.80 &	2.01 &	0.37 &	31.21\\
15.0 &	1.70 &	2.60 &	1.85 &	0.34	& 30.66\\
    \bottomrule
    \end{tabular}
    }
    \caption{Latency results in two simulated load tests on L2 ranking (plus back end API). Query-per-second (QPS) is the average request load in the test. Back end API takes about 20 ms for each query. Most queries can be processed within 1-2 second, even with relatively high request load.}
    \label{tbl:L2-test}
    \vspace{-5mm}
\end{table}

For L2, we used Kubernetes to manage a GPU cluster. See \autoref{fig:system} (right) for a reference architecture. 
We used V100 GPUs in initial experiments. Since they are relatively expensive, we explored using low-cost GPUs in subsequent experiments to maximize cost effectiveness. 
For each query, to rerank the top 60 candidate paragraphs from L1, it takes about 0.9 second on a V100 GPU. 
The K80 GPU only costs a fraction of V100, but requires 3 second per query. We therefore used 4-K80 machines, which reduce the latency to 0.75 second but cost less than a third of the cost for V100.

\autoref{tbl:L1-test} and \autoref{tbl:L2-test} shows simulated test results for L1 and L2 ranking, respectively. There were no failures in all the tests. For L1 ranking, our configuration can already support 10-20 QPS while keeping latency for most queries to less than a second. To support higher QPS, we can simply add more main and data nodes, which scale roughly linearly. For L2 ranking, our test used 32 4-K80 machines with a total of 128 K80s. It can support about 10 QPS while keeping latency for most queries to around or under a second. To support higher QPS, we can simply add more K80 machines.
 
\subsection{Microsoft Biomedical Search}

Our biomedical search system has been deployed as \DNAME, which is publicly available. See \autoref{fig:ux} for a sample screenshot. 

Before deployment, we conducted several user studies among the co-authors and our extended teams with a diverse set of self-constructed and sampled queries. Overall, we verified that our system performed well for long queries with complex intent, generally returning more relevant results compared to PubMed and other search tools. However, for overly general short queries (e.g., ``breast cancer''), our system can be under-selective among articles that all mention the terms. To improve user experience, we augmented L1 ranking by including results from Microsoft Academic, which uses a \emph{saliency} score that takes into account temporal evolution and heterogeneity of the Microsoft Academic Graph to predict and up-weight influential papers~\cite{wang2019ma-paper, wang2020microsoft, sinha2015ma-overview}.
Given a query, we retrieve top 30 results from Microsoft Academic as ranked by its saliency score and combine them with the top 30 results from BM25. L2 reranking is then conducted over the combined set of results. The saliency score helps elevate important papers when the query is underspecified, which generally leads to a better user experience.

In addition to standard search capabilities, our system incorporates a state-of-the-art machine reading comprehension (MRC) method \cite{cheng2021unitedqa} trained on \cite{kwiatkowski2019nq} as an optional component. Given a query and a top reranked candidate passage, the MRC component will treat it as a question-answering problem and return a text span in the passage as the answer, if the answer confidence is above the score of abstaining from answering. 
The MRC component uses the same cloud architecture as the L2 neural ranker \autoref{fig:system} (right), with similar latency performance. 

\eat{
\begin{table}[t!]
    \centering
    
    \setlength{\tabcolsep}{1.0mm}{
    \begin{tabular}{lcccc}
    \toprule
Search Engine & NDCG@10 & P@5   \\
\midrule
BM25 & 32.6 &  29.5 \\
SciBERT & 40.3 &  34.3 \\
PubMed & 31.2 & 26.7 \\
\bottomrule
    \end{tabular}
    }
    \caption{Evaluation Results of Cleveland Queries}
    \label{}
    \vspace{-5mm}
\end{table}
}



\begin{table}[t!]
    \centering
    \setlength{\tabcolsep}{1.0mm}{
    \begin{tabular}{ccccc}
    \toprule
QPS & L1 (Cost) & L2 (Cost) & MRC (Cost) & Total Cost  \\
\midrule
4 &	13 D8v3 (\$5K) &	32 K80 (\$10K) &	48 K80 (\$14K) &	\$29K\\
7 &	13 D8v3 (\$5K) &	64 K80 (\$20K) &	96 K80 (\$28K) &	\$53K\\
14 &	13 D8v3 (\$5K) &	128 K80 (\$40K) &	192 K80 (\$55K) &	\$100K\\
28 &	26 D8v3 (\$10K) &	256 K80 (\$80K) &	384 K80 (\$110K) &	\$200K\\
    \bottomrule
    \end{tabular}
    }
    \caption{Reference configuration and monthly cost estimate 
    to support expected QPS while keeping median latency under two seconds (based on pricing from June 2021).
    }
    \label{tbl:system-cost}
    \vspace{-3mm}
\end{table}


Our system can be deployed for public release at a rather affordable cost. \autoref{tbl:system-cost} shows the reference configuration and cost estimate to support various expected loads (QPS).

\eat{
--------------------------------------------
The whole Biomed Search system can be divided into following components:
1.	Web front end
2.	Web Back end API
3.	Cache
4.	L2 reranking layer (Host on Kubernetes cluster)
5.	L1 retrieval layer (Elastic Search)

One query is sent from Web front end to Web back end API. The back end API coordinates the L1 and L2 layer. It first consults the cache layer, returns results directly if the query exist in cache. If the query could not fulfil by cache, back end API retrieves top 60 results from back end and then re-rank them with L2 layer. The final result list are deduplicated. 
 
We adopt the two stage ranking methodology to reduce the computational cost of the system.
 
The key challenge to the system is to ensure scalability. We build the whole system on top of Azure. The Web front end, Web back end and cache part are relatively fast compared to L1 & L2 layers, so we will focus on the L1 & L2 scalability.
 
At L1 layer, we inject the whole PubMed corpus. The whole corpus contains 30M entries, 20M (19628873) of them has abstract and 3M (3008541) papers have full text. For each entry, we divide the raw text into paragraphs to allow retrieval at finer granularity, but the decision introduce further challenge to the system's scalability. The final paragraph number is 216M (216226842). With such volume of corpus, the index size of Elastic Search is 164.7GB. The size would further grows linearly with the additional replication we added to ensure system availability.  According to our measurement, the search latency of Elastic Search is usually linear or slightly sub-linear to the corpus size. Therefore, before fine tuning, the L1 layer experience long latency.
 
At L2 layer, although L2 model is only ran against limit number of query & paragraph pairs. The L2 model itself is large scale pretrained NLP model like BERT, RoBERTa, which are known to be computationally heavy. Therefore, we need a cluster of GPUs as well as careful hardware/software tuning to maximize the ROI which minimize the L2 latency.
 
We have run several rounds of experiments to measure the system performance, including L1 layer only, Back end API + L1 layer, Back end API + L2 layer. Each experiment adds only one component, which helps us to identify the bottle neck.
The back end API layer has very little impact to the overall latency when tuned appropriately, so in this paper, we will only introduce the last 2 rounds of experiments.
Below we introduce the methodology of the experiment, the 2 rounds of experiments used the same methodology.
•	Tools : Locust, which is a python modern load testing framework.
•	Hardware/Software configuration
o	The back end API is developed with Flask, use Gevent + 8 workers to ensure highest performance
o	Deploy the back end API, L1 layer and L2 layer in the same data center to minimize the network cost. The machine to send queries are also deployed in the same data center.
o	All the server are deployed in the same virtual network.
o	The final cost of back end API is around 20ms.
•	Query set
o	We prepare a query set which contains 71K queries sampled from MA query log.
•	Experiment setting
o	Locust client is able to simulate many users. To simulate the real user experience, we allow each user to randomly wait around 15s to 60s  after each search. Each experiments runs for 10 minutes.
o	We turned off cache layer during all experiments.
o	Elastic Search might cache results to speed up repeat queries. To be strict, we leverage our large 71K queries set, to ensure there are no repeat query in each experiment. However, we find Elastic Search need a little bit warmup before hitting its maximum capacity, so we warm up the system with low QPS (0.5 per sec) for 10 minutes before true experiments. According to our experience, after warm up, the cold query (new query) and hot query (repeat query) has around 10
 
For the L1 layer, based on the performance test, we choose following Elastic Search hardware/software configuration.
•	10 premium machines (D8s v3) as data nodes. each machine attaches 1 p30 SSD disk (p30 means 1TB). 
•	3 premium machines as master nodes.
•	use 30 shards
 
The query is sent to master node. The master node distribute the query to each data node and then gather and merge the results. Data node is mainly IO intensive, master node is more computation intensive. 
 
Below are the test result under such setting.
 
# Fails	RPS	Median (ms)	90
0	13.2409 	510	750	592	225	7072
0	26.7757 	 600	1500	876	222	31046
 
For the L2 layer, we find single machine is not capable to handle high cost. That is why we choose to use a cluster, all the nodes are managed by Kubernetes. The architecture is displayed in P32 of slides.
 
At the beginning, we choose 1 V100 GPU on each node, which costs 0.9s for a typical query. To maximize the ROI, we also test configuration of 1 K80S GPU (3s) and the n use 4 K80s GPU to handle one query (with multiple paragraphs), which costs 0.75s. Our later test are conducted by scale out to 128 K80S GPUs.
 
The scalability of L2 is linear to the number of GPUs.
 
RPS	Median (ms)	90
14.785	1800	2800	2010	367	31205
15	1700	2600	1849	335	30659
 
\begin{table}[]
\begin{tabular}{|l|l|l|l|}
\toprule
\hline
COVID-19 search   engine                   & URL                                                   & Dataset         & Retrieval \&   ranking method                                                                                                                                                                                                                        \\ \hline
\midrule
Covidex (U of   Waterloo, NYU)             & https://covidex.ai/                                   & CORD-19         & Retrieval:   BM25 + T5 based doc2query doc expansion, Ranking: T5 based reranker https://arxiv.org/abs/2007.07846                                                                                                                                    \\ \hline
Azure COVID19   search  (Microsoft Azure)  & https://covid19search.azurewebsites.net/home/index?q= & CORD-19         & Azure Cognitive   Search: BM25 according to Azure Cognitive Search                                                                                                                                                                                   \\ \hline
COVID-19   Research Explorer (from Google) & https://covid19-research-explorer.appspot.com/        & CORD-19         & Term   based + encode query \& doc with BERT and retrieve by KNN https://ai.googleblog.com/2020/05/an-nlu-powered-tool-to-explore-covid-19.html                                                                                                      \\ \hline
COVID-19   Search (from SalesForce)        & https://sfr-med.com/search                            & CORD-19         & Siamese   BERT https://www.salesforce.com/news/stories/salesforce-research-develops-new-search-engine-to-support-the-fight-against-covid-19/                                                                                                         \\ \hline
CORD-19   Explorer (from AI2)              & https://cord-19.apps.allenai.org/                     & CORD-19         & Semantic   Scholar technique: Retrieval: Elastic Search + Reranker: LightGBM ranker with   a LambdaRank objective  https://github.com/allenai/s2search https://medium.com/ai2-blog/building-a-better-search-engine-for-semantic-scholar-ea23a0b661e7 \\ \hline
Primer.ai                                  & https://covid19primer.com/                            & CORD-19 \& news & https://covid19primer.com/About It employs natural language processing   and generation to read and analyze research papers.                                                                                                                         \\ \hline
\bottomrule
\end{tabular}
\end{table}

}

%% file: 5_discussion.tex
\section{Discussion}


Prior work on vertical search tends to focus on domain-specific crawling (focused crawling) and user interface~\cite{baeza2011modern}. 
We instead explore the orthogonal aspect of the underlying search algorithm. These tend to be simplistic in past systems, due to the scarcity of domain-specific relevance labels, as exemplified by the PubMed search engine. 
While easier to implement and scale, such systems often render subpar search experiences, which is particularly concerning for high-value verticals such as biomedicine. E.g., \citet{soni2021evaluation} studied the evaluation of commercial COVID-19 search systems and found that ``{\em commercial search engines sizably underperformed those evaluated under TREC-COVID. This has implications for trust in popular health search engines and developing biomedical search engines for future health crises.}'' 

By leveraging domain-specific pretraining and self-supervision from broad-coverage query-passage dataset, we show that it is possible to train a sophisticated neural ranking system to attain high search relevance, without requiring any manual annotation effort. 
Although we focus on biomedical search as a running example in this paper, our reference system comprises general and reusable components that can be directly applied to other domains. 
Our approach may potentially help bridge the performance gap in conventional vertical search systems while keeping the design and engineering effort simple and affordable.

There are many exciting directions to explore. For example, we can combine our approach with other search engines that take advantage of complementary signals not used in ours. Our hybrid L1 ranker combining BM25 with Microsoft Academic Search saliency scores is an example of such fusion opportunities. A particularly exciting prospect is applying our approach to help improve the PubMed search engine, which is an essential resource for millions of biomedical practitioners across the globe. 

In the long run, we can also envision applying our approach to other high-value domains such as finance, law, retail, etc. Our approach can also be applied to enterprise search scenarios, to facilitate search across proprietary document collections, which standard search engines are not optimized for. 
In principle, all it takes is gathering unlabeled text in the given domain to support domain-specific pretraining. If a comprehensive index is not available (as in PubMed for biomedicine), one could leverage focused crawling in traditional vertical search to identify such in-domain documents from the web. 
In practice, additional challenges may arise, e.g., in self-supervised fine-tuning. Currently, we generate the training dataset by selecting MARCO queries using a domain lexicon. If such a lexicon is not readily available (as in UMLS for biomedicine), additional work is required to identify words most pertinent to the given domain (e.g., by contrasting between general and domain-specific language models). We also rely on MARCO to have sufficient coverage for a given domain. We expect that high-value domains are generally well represented in MARCO already. For an obscure domain with little representation in open-domain query log, we can fall back to using a general query-document relevance model as a start and invest additional effort for refinement. 

%% file: 6_conclusion.tex
\vspace{-5mm}
\section{Conclusion}

We described a methodology for developing vertical search capabilities and demonstrate its effectiveness in the TREC-COVID evaluation for COVID-related biomedical search. 
The generality and efficacy of the approach rely on domain-specific pretraining and self-supervised fine-tuning, which require no annotation effort for applying to a new domain. 
Using biomedicine as a running example, we present a general reference system design that can scale to tens of millions of domain-specific documents by leveraging capabilities supplied in modern cloud infrastructure. Our system has been deployed as Microsoft Biomedical Search. Future directions include further improvement of  self-supervised reranking, combining the core retrieval and ranking services with complementary search methods and resources, and validation of the generality of the methodology by testing the approach in building search systems for other vertical domains. 